\documentclass{article}
\usepackage{microtype}
\usepackage{amsmath,amsfonts}
\usepackage{array}
\usepackage{textcomp}
\usepackage{url}
\usepackage{verbatim}
\usepackage{graphicx}
\usepackage{booktabs}
\usepackage{multirow}
\usepackage{xcolor}
\usepackage{colortbl}

\usepackage{siunitx}
\DeclareSIUnit\mac{MAC}
\DeclareSIUnit\mmac{MMAC}
\DeclareSIUnit\cycle{cycle}
\DeclareSIUnit\fps{FPS}

\newcommand{\rowrule}{\arrayrulecolor{black!18}\specialrule{0.3pt}{0.35ex}{0.35ex}\arrayrulecolor{black}}

\newcommand{\devsq}[1]{\textcolor{#1}{\rule{0.6em}{0.6em}}\,}
\newcommand{\typesq}[1]{\tikz[baseline=0pt]{\draw[line width=0.3pt, fill=#1]
  (0,0) rectangle (0.6em,0.6em);}\,}
\newcommand{\cpusq}{\typesq{white}}
\newcommand{\npusq}{\typesq{black!45}}
\newcommand{\dspsq}{\typesq{black}}
\definecolor{mobilenetcolor}{HTML}{234E9A}
\definecolor{mcunetcolor}{HTML}{4AA0D8}
\definecolor{efficientnetcolor}{HTML}{127A5A}
\definecolor{squeezenetcolor}{HTML}{7DBF55}

\usepackage{tikz}
\usetikzlibrary{positioning, arrows.meta, fit, backgrounds, calc}

\usepackage{hyperref}

\usepackage[accepted]{mlsys2025}

\makeatletter
\renewcommand{\Notice@String}{}
\makeatother

\makeatletter
\long\def\@makecaption#1#2{
 \vskip 4pt
        \baselineskip 11pt
        \setbox\@tempboxa\hbox{#1. #2}
        \ifdim \wd\@tempboxa >\hsize
        \sbox{\newcaptionbox}{\small\sl #1.~}
        \newcaptionboxwid=\wd\newcaptionbox
        \usebox\newcaptionbox {\footnotesize #2}
        \else
          \centerline{{\small\sl #1.} {\small #2}}
        \fi}
\makeatother

\setcitestyle{numbers,square,citesep={,}}

\usepackage[nohyperlinks]{acronym}
\acrodef{AI}{artificial intelligence}
\acrodef{CMOS}{complementary metal-oxide-semiconductor}
\acrodef{CNN}{convolutional neural network}
\acrodef{CPU}{central processing unit}
\acrodef{DMA}{direct memory access}
\acrodef{DNN}{deep neural network}
\acrodef{EEG}{electroencephalography}
\acrodef{DSP}{digital signal processor}
\acrodef{EMG}{electromyography}
\acrodef{HWC}{hardware convolution engine}
\acrodef{IoT}{Internet of Things}
\acrodef{MCU}{microcontroller unit}
\acrodef{NAS}{neural architecture search}
\acrodef{NPU}{neural processing unit}
\acrodef{PRISMA}{Preferred Reporting Items for Systematic Reviews and Meta-Analyses}
\acrodef{PULP}{Parallel Ultra-Low Power}
\acrodef{SIMD}{single instruction, multiple data}
\acrodef{SoC}{system-on-chip}
\acrodef{SPOT}{Subthreshold Power Optimized Technology}
\acrodef{SRAM}{static random-access memory}
\acrodef{TinyML}{tiny machine learning}

\graphicspath{{figures/}}
\mlsystitlerunning{TinyML Hits the Memory Wall: A Benchmark of Ten Sub-Watt Edge Devices}

\begin{document}

\twocolumn[
\mlsystitle{TinyML Hits the Memory Wall: A Benchmark of Ten Sub-Watt Edge Devices} 

\begin{mlsysauthorlist}
\mlsysauthor{Pietro Bonazzi}{eth}
\mlsyscorrespondingauthor{Pietro Bonazzi}{pbonazzi@ethz.ch}
\mlsysauthor{Luigi Capogrosso}{itu}
\mlsysauthor{Ahmet Celik}{eth}
\mlsysauthor{Rosanna Manzo}{unina}
\mlsysauthor{Marco Giordano}{eth}
\mlsysauthor{Szymon Ruci\'{n}ski}{eth,csem}
\mlsysauthor{Jonathan Hungerbuehler}{eth}
\mlsysauthor{Michele Magno}{eth}
\end{mlsysauthorlist}

\mlsysaffiliation{eth}{ETH Z\"{u}rich, Z\"{u}rich, Switzerland}
\mlsysaffiliation{itu}{Interdisciplinary Transformation University of Austria}
\mlsysaffiliation{unina}{University of Naples Federico II, Naples, Italy}
\mlsysaffiliation{csem}{CSEM, Neuch\^{a}tel, Switzerland}

\mlsyskeywords{edge AI, TinyML, microcontroller, MCU, hardware-software co-design, benchmark, energy efficiency,
convolutional neural network, inference latency, neural processing unit}

\vskip 0.3in

\begin{abstract}
Neural network inference at the edge must fit tight energy budgets.
However, existing benchmarks compare a small set of devices, do not report energy on a common basis, and do not include the newest microcontrollers with an \acs{NPU} or the \qty{1}{\giga\hertz} Cortex-M85 microcontrollers.
In addition, hardware-aware model design ranks networks by MAC count, although no study has tested whether this proxy holds across accelerators.
To fill these gaps, we benchmark sub-watt edge \ac{AI} hardware. Specifically, we evaluate four convolutional neural network models (MobileNetV3, MCUNetV1, EfficientNet, and SqueezeNet) and two tiny transformer networks (MobileViT-XXS and DeiT-tiny) across ten commercially available edge devices: an in-sensor \ac{DSP} (Sony IMX500), four \acsp{MCU} with an \acs{NPU} (STMicroelectronics STM32N6, Alif Ensemble E7/E8 and GreenWaves GAP9), and five \acsp{MCU} without an \acs{NPU} (STMicroelectronics STM32U5, Ambiq Apollo510B, Espressif ESP32-S3 with its Xtensa LX7 core, Renesas RA8M2, and the recently released STMicroelectronics STM32V8).
Four of them are novel devices: the Alif Ensemble E8, Ambiq Apollo510B, Renesas RA8M2 and STM32V8.
Measured as energy above idle, the Sony IMX500 (\qtyrange{0.13}{0.35}{\milli\joule}) uses 1.3--3$\times$ less energy than the STM32N6 (\qtyrange{0.31}{1.05}{\milli\joule}) on every model, the GAP9 (\qtyrange{0.11}{1.23}{\milli\joule}) uses the least on MobileNetV3, and the Alif E7 (\qtyrange{0.12}{0.45}{\milli\joule}) uses the least on MCUNetV1 and SqueezeNet; all four use 6--34$\times$ less energy than the STM32V8 \acs{MCU} (\qtyrange{1.85}{3.98}{\milli\joule}), and the Alif E7 and IMX500 form the power--performance Pareto front.
Moreover, correlating measured latency with each model's memory traffic shows that all four accelerators running the \acsp{CNN} are memory-bound (latency scales with the bytes of weights and activations a model must read and write, $r\ge0.98$, and not with its MACs), whereas no \acs{MCU} without an \acs{NPU} is; the \acs{NPU} advantage scales with arithmetic intensity, from $1.4\times$ over a \qty{1}{\giga\hertz} Cortex-M85 on MCUNetV1 to $8.2\times$ on the high-intensity SqueezeNet. Consequently, MAC count is a misleading cost proxy for sub-watt accelerators.
\end{abstract}

]

\printAffiliationsAndNotice{}
\acresetall

\section{Introduction} \label{sec:intro}
\begin{figure}[!t]
  \centering
  \includegraphics[width=0.8\columnwidth]{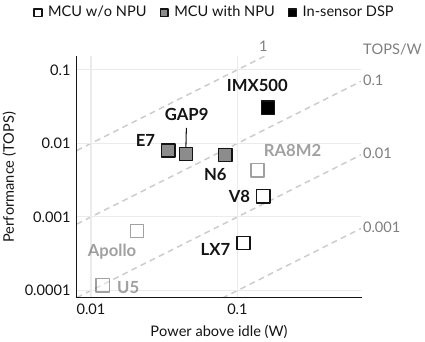}
  \caption{Measured performance vs. power above idle during inference of the six devices with measured energy, each the geometric mean over the four \acp{CNN} (three on the STM32V8). Faded, indicative only and at total power: Apollo510B as vendor-reported on a different model (\qty{20.7}{\milli\watt}, FOMO~\citep{ambiqfomo}), STM32U5 and RA8M2 from their datasheets.}
  \label{fig:intro}
\end{figure}

Vision on sub-watt \acp{MCU} and smart sensors faces tight budgets: a camera node or wearable must classify an image in milliseconds, with millijoules of energy and megabytes of memory~\citep{Warden2019}.
Meanwhile, three new classes of silicon target this regime:
\acp{MCU} now integrate dedicated \acp{NPU} (Arm Ethos-U55 and U85, ST Neural-ART), image sensors run networks next to the pixel array, and \acsp{MCU} built around the Arm Cortex-M85 feature Helium vector extensions reaching \qty{800}{\mega\hertz} to \qty{1}{\giga\hertz}~\citep{stm32v8, ra8m2} clock frequency.
A designer choosing among them therefore faces one question: when is a dedicated accelerator worth the cost, and for which models?
Answering this question requires a benchmark that existing studies do not provide, for three reasons.
First, they cover a small set of devices, each from one architecture class or an earlier device generation~\citep{Lai2018, Millar2025, Capogrosso2026c}. For example, MLPerf Tiny ~\citep{mlperftinyv14, mlperftinyresults} spans many vendors, however includes none of our ten devices except the STM32U5; infact to the best of our knowledge, four of the ten devices we study have no published inference benchmark at all.
Second, energy is not reported on a common basis: some benchmarks omit it~\citep{Lai2018}, and the others measure it on device sets that do not span \acsp{MCU}, \acsp{MCU} with an \acs{NPU} and in-sensor \acsp{DSP}~\citep{Banbury2021, Millar2025, Capogrosso2026c}, so no study compares the energy of these classes on the same models.
Third, hardware-aware model design ranks networks by MAC count, or by latency on one target, and assumes that this ranking transfers across devices~\citep{Howard2019, Lin2020}. No study has measured whether on current accelerators execution is limited by arithmetic or by data movement~\citep{Williams2009}.

To provide this benchmark, we measure ten commercial edge devices spanning five architecture classes: an in-sensor \ac{DSP} (Sony IMX500), \ac{MCU}-integrated \acsp{NPU} (STM32N6, Alif Ensemble~E7 and E8), a RISC-V cluster with a convolution accelerator (GreenWaves GAP9), high-clock Cortex-M85 \acsp{CPU} (STM32V8, Renesas RA8M2), and low-power Cortex-M and Xtensa cores (STM32U5, Ambiq Apollo510B, Espressif ESP32-S3).
Four \acp{CNN} run on the nine devices other than the Alif E8, wherever memory and toolchain allow, and two vision transformers run on the devices built for attention.
Every model is deployed with the vendor's toolchain and measured on the device with one protocol. The models are chosen so that their MACs and memory traffic vary independently (Section~\ref{subsec:models}).
To make energy comparable, we report the energy above idle, which removes the static power of everything outside the computation.

\textit{\textbf{What we find.}}
All four accelerators that run the \acp{CNN} (the IMX500 and three \acsp{NPU}) are memory-bound: their latency grows with the total memory traffic $Q$, the bytes of weights and activations that the model's layers must read and write ($r_Q\ge0.98$), and not with its MACs (Fig.~\ref{fig:bound} and Table~\ref{tab:bound}), whereas no \acs{MCU} without an \acs{NPU} is.
MAC count therefore misranks models on these accelerators: SqueezeNet performs $8.5\times$ the MACs of MobileNetV3 yet runs $1.8\times$ faster on the Alif Ensemble~E7, because it requires less memory (1.22 versus \qty{1.76}{\mega\byte}).
For the same reason, the \ac{NPU} advantage over a \qty{1}{\giga\hertz} Cortex-M85 depends on arithmetic intensity, from $1.4\times$ on MCUNetV1 to $8.2\times$ on the high-intensity SqueezeNet.
Finally, measured as energy above idle, the IMX500, GAP9, Alif E7 and STM32N6 use 6--34$\times$ less energy per inference than the STM32V8 \acs{MCU}.
Averaged over the \acp{CNN} suite and on power above idle, the Alif E7 (8.0~GOPS at \qty{34}{\milli\watt}) and the IMX500 (31~GOPS at \qty{162}{\milli\watt}) form the power--throughput Pareto front (Fig.~\ref{fig:intro}), with the GAP9 (7.2~GOPS at \qty{44}{\milli\watt}) just behind; the STM32N6 matches the GAP9's throughput at $1.9\times$ its power.

In summary, our contributions are as follows.
\begin{enumerate}
  \item \textit{A bound analysis}: by checking whether each device's latency follows a model's compute (MACs) or its data movement (the bytes of weights and activations it reads and writes), we show that sub-watt accelerators are memory-bound while \acsp{MCU} without an \acs{NPU} are not, and turn this into guidance on when an \ac{NPU} is worth it and what hardware-aware design should optimize.
  \item \textit{The broadest measured comparison of sub-watt vision hardware to date:} ten devices and five architecture classes on identical models and inputs, reporting latency, energy, MAC/cycle, B/cycle and MMAC/J.
  \item \textit{Measurements on novel devices:} the Alif Ensemble~E8, Ambiq Apollo510B, Renesas RA8M2 and STM32V8, including vision transformers on the Ethos-U85 of the Alif E8.
\end{enumerate}

\section{Related Work} \label{sec:related}

\textit{\textbf{Device coverage.}}
MLPerf Tiny~\citep{Banbury2021} defines a benchmark suite for microcontroller-class hardware on the MLPerf inference methodology~\citep{Reddi2020}, but its closed division fixes one reference model per task and it covers only the devices for which results are submitted, whereas we choose models whose MACs and memory traffic vary independently and measure every device ourselves.
Lai~\textit{et al.}~\citep{Lai2018} show that CMSIS-NN kernels reach near-peak utilization of Cortex-M arithmetic units; they study one class of \ac{CPU}, whereas we compare such \acsp{CPU} with \acsp{NPU} and an in-sensor \ac{DSP}.
Millar~\textit{et al.}~\citep{Millar2025} compare \acsp{MCU} with an \ac{NPU} (Analog Devices MAX78000, Himax HX-WE2, NXP MCXN947, GreenWaves GAP8 and the CV1800B) against \acsp{MCU} without one (STM32H7, ESP32-S3) under a common compilation pipeline, and find the latter less efficient on every model; their devices belong to an earlier generation, whereas we add the Ethos-U85, Neural-ART, an in-sensor accelerator and the \qty{1}{\giga\hertz} Cortex-M85, deploy each with its vendor's toolchain, and explain the gap by whether a device is limited by compute or by data movement.
Closest to our work, Capogrosso~\textit{et al.}~\citep{Capogrosso2026c} compare edge platforms on a single workload and find that the IMX500 achieves the highest MAC/cycle and the lowest energy-delay product; we extend the comparison to six models and ten devices, which shows that this MAC/cycle depends on the model's arithmetic intensity.
Surveys of machine learning on \acp{MCU}~\citep{Saha2022, LucanOrasan2022, Warden2019} and of deep learning and RISC-V accelerators~\citep{Silvano2025, SanchezFlores2022, Kalapothas2023, Agosta2025, Flamand2018} describe architectures, kernels and toolchains, but do not measure devices against each other, which is what we do.
Six works automate \ac{DNN} deployment on accelerator-equipped \acp{SoC}~\citep{Macan2023, Mirsalari2024, Mazzoni2024, AmineHamdi2025, Jung2025, Mirsalari2025}, because the compiler can matter more than the peak throughput of the silicon; each targets its own platform, whereas we compare commercial devices as a practitioner deploys them, with each vendor's toolchain.

\textit{\textbf{Energy on a common basis.}}
Lai~\textit{et al.}~\citep{Lai2018} do not report energy, and Millar~\textit{et al.}~\citep{Millar2025} and Capogrosso~\textit{et al.}~\citep{Capogrosso2026c} report it on the supply each board exposes, which includes a different amount of static power on each device, so their raw energy per inference is not comparable across devices.
We instead compare devices on the energy above idle, which removes this static power (Section~\ref{subsec:methodology}).

\textit{\textbf{MAC count as a cost proxy.}}
Hardware-aware model design has produced the compact networks we benchmark, but each was designed for one kind of target, whereas we measure them on ten.
Depthwise separable convolutions~\citep{Howard2017} reduce MACs on the assumption that MACs set the cost; we show that on sub-watt accelerators they do not.
MobileNetV3~\citep{Howard2019} includes measured latency on a mobile \ac{CPU} in its \ac{NAS} objective, and MCUNet~\citep{Lin2020} co-designs the network with its inference engine for the \ac{SRAM} of one \ac{MCU}; we test whether the resulting models keep their advantage on devices they were not designed for, such as every device in our benchmark.
Both are instances of hardware--software co-design~\citep{Capogrosso2024}, which needs to know what limits each device; we provide this by measuring whether execution is limited by arithmetic or by data movement, which no prior work has measured on current accelerators.
To do so we use models whose MACs and memory traffic vary independently, and add two vision transformers~\citep{Touvron2021, Mehta2022} for the devices that support attention.

\section{Benchmark} \label{sec:benchmark}

The benchmark is built on three principles.
\textit{i) Identical workloads:} every device runs the same models at the same input resolution, so that differences reflect the hardware and its toolchain rather than the workload.
\textit{ii) Deployment as a practitioner would do it:} each model is quantized to INT8 and compiled with the vendor's recommended toolchain, and a model that does not fit a device, or uses operators its toolchain does not support, is reported as missing rather than modified.
\textit{iii) Measurement on the device:} latency and energy are never taken from datasheets or simulators.

\subsection{Hardware Platforms} \label{subsec:hardware}

\begin{table}[!t]
  \centering
  \caption{Evaluated devices. Core: Arm Cortex-M unless noted. Clock: of the unit that runs
           the model (the \ac{NPU} where present). SRAM of the Xtensa LX7: internal SRAM + external PSRAM.}
  \label{tab:hardware}
  \resizebox{\columnwidth}{!}{%
  \begin{tabular}{@{}l l l l r r@{}}
    \toprule
    \textbf{Device} & \textbf{Core} & \textbf{SIMD} & \textbf{Accelerator} & \begin{tabular}[b]{@{}c@{}}\textbf{Clock}\\\textbf{(MHz)}\end{tabular} & \begin{tabular}[b]{@{}c@{}}\textbf{SRAM}\\\textbf{(MB)}\end{tabular} \\
    \midrule
    \cpusq STM32U5~\citep{stm32u5} & M33 & 32-bit  & --& 160 & 3.0 \\
    \rowrule
    \cpusq Xtensa LX7~\citep{esp32s3} & LX7 & 128-bit  & --& 240 & 0.5\,+\,8 \\
    \rowrule
    \cpusq STM32V8~\citep{stm32v8} & M85 & Helium  & --& 800 & 1.5 \\
    \rowrule
    \cpusq Apollo510B~\citep{apollo510b} & M55 & Helium  & --& 250 & 3.0 \\
    \rowrule
    \cpusq RA8M2~\citep{ra8m2} & M85 & Helium  & --& 1000 & 2.0 \\
    \rowrule
    \npusq STM32N6~\citep{Desoli2026} & M55 & Helium  & Neural-ART& 1000 & 4.2 \\
    \rowrule
    \npusq Alif E7~\citep{alife7} & M55 & Helium  & Ethos-U55& 400 & 13.5 \\
    \rowrule
    \npusq Alif E8~\citep{alife8} & M55 & Helium  & Ethos-U85& 400 & 9.8 \\
    \rowrule
    \npusq GAP9~\citep{Flamand2018} & RISC-V & packed  & NE16& 370 & 2.1 \\
    \rowrule
    \dspsq IMX500~\citep{imx500} & -- & --  & DSP& 262.5 & 8.0 \\
    \bottomrule
  \end{tabular}%
  }
\end{table}

We evaluate ten commercial edge \ac{AI} devices that span five architecture classes and clock rates from 160 to \qty{1000}{\mega\hertz} (Table~\ref{tab:hardware}): an in-sensor \acs{DSP} (Sony IMX500), \ac{MCU}-integrated \acsp{NPU} (STM32N6, Alif E7 and Alif E8), a RISC-V cluster with a convolution accelerator (GAP9), high-clock Cortex-M85 \acsp{CPU} with Helium vector extensions (STM32V8, Renesas RA8M2), and low-power \acsp{MCU} without an \acs{NPU}: the Apollo510B (Cortex-M55 with Helium), the STM32U5 (Cortex-M33 with 32-bit \ac{SIMD} instructions only) and the Xtensa~LX7 of the ESP32-S3 (128-bit \ac{SIMD}).
We chose them to cover the three design options available to a sub-watt vision system today: computing next to the pixels, adding an accelerator to an \ac{MCU}, or relying on a fast vector \ac{CPU}.

\textit{\textbf{Alif Ensemble E7}}~\citep{alife7}.
The Alif E7 features dual Arm Cortex-M55 cores with Helium vector extensions and an Ethos-U55 \ac{NPU} and a dedicated \ac{SRAM} for activations.
Apart from vendor figures, its only published latency and energy evaluation is a recent preprint by \citet{fanariotis2025evaluating}.

\textit{\textbf{Alif Ensemble E8}}~\citep{alife8}.
The Alif E8, announced in 2025, belongs to the second Ensemble generation and targets transformer workloads: alongside two Cortex-M55 cores with Helium (a \qty{400}{\mega\hertz} high-performance and a \qty{160}{\mega\hertz} high-efficiency core), two Cortex-A32 application cores and two Ethos-U55 \acsp{NPU}, it adds an Arm Ethos-U85 \ac{NPU} with native support for transformer operators, and has \qty{9.75}{\mega\byte} of on-chip \ac{SRAM} and \qty{5.5}{\mega\byte} of MRAM.
Because the Alif E8 targets transformers, we evaluate it on two vision transformers rather than on the four \acp{CNN} (Section~\ref{subsec:latency}). The Ethos-U85 and the high-performance Cortex-M55 run at \qty{400}{\mega\hertz} by default, the clock we use for MAC/cycle.

\textit{\textbf{Ambiq Apollo510B}}~\citep{apollo510b}.
The Ambiq Apollo510B combines a \qty{250}{\mega\hertz} Cortex-M55 core with Helium vector extensions with the company's \ac{SPOT} platform for low-voltage operation; it has no \ac{NPU}, so inference runs on the Helium datapath of the \ac{CPU}.

\textit{\textbf{Espressif ESP32-S3 (Xtensa LX7)}}~\citep{esp32s3}.
The ESP32-S3 integrates a dual-core Xtensa LX7 processor with 128-bit \ac{SIMD} vector extensions at \qty{240}{\mega\hertz}, \qty{512}{\kilo\byte} of internal \ac{SRAM} and, on our board, \qty{8}{\mega\byte} of PSRAM, making it the \acs{MCU} without an \acs{NPU} with the largest (external) memory.
Owing to its large installed base and low cost, it has been used for \ac{TinyML} deployments in human activity recognition~\citep{Pfitzinger2023}, speech and sound classification~\citep{Wu2024, Kristiawan2025, Chowdhury2026}, time-series forecasting~\citep{Moura2026, Na2026}, \acs{SIMD}-accelerated inference optimization~\citep{Zhang2025}, split inference across cores~\citep{Palangsantikul2025}, and \ac{IoT} monitoring~\citep{Walter2024, Manea2025, Chen2025, PavanKumarS2026, Mondal2026, Song2026, Le2026, Elumalai2026, Karasu2026}.

\textit{\textbf{GreenWaves GAP9}}~\citep{Flamand2018}.
The GreenWaves GAP9 is a \qty{22}{\nano\metre} RISC-V \ac{SoC} comprising a single-core fabric controller and an eight-core compute cluster, whose RISC-V cores support packed 8- and 16-bit \ac{SIMD} instructions, augmented by the NE16 \ac{HWC}.
Its \ac{PULP} architecture enables data-parallel execution.
GreenWaves Technologies entered judicial liquidation in January 2025~\citep{greenwavesliquidation}, so GAP9 also illustrates the supply risk of building on a start-up's silicon.
Among the ten devices, GAP9 is the most studied, with published wearable and embedded \ac{AI} applications in biosignal processing for seizure detection and \ac{EEG}-based interfaces~\citep{Ingolfsson2023, Busia2024, Ingolfsson2024, Ingolfsson2025}, \ac{EMG}-driven gesture and hand dynamics estimation~\citep{Zanghieri2023, Zanghieri2024a, Zanghieri2024b}, drowsiness and eye-movement classification~\citep{Frey2024}, object detection and multi-object tracking~\citep{Moosmann2024, Boyle2024, Bompani2024a, Bompani2024b, Bompani2026}, on-device continual and self-supervised learning~\citep{Cereda2024a, Cereda2024b, Wibowo2024, Mei2024, Mei2025, Nadalini2026}, and automated \ac{DNN} compilation and deployment~\citep{Macan2023, Mirsalari2024, Mazzoni2024, AmineHamdi2025, Jung2025, Mirsalari2025}.

\textit{\textbf{Renesas RA8M2}}~\citep{ra8m2}.
The Renesas~RA8M2 pairs a \qty{1}{\giga\hertz} Cortex-M85 core with Arm Helium vector extensions and a \qty{250}{\mega\hertz} Cortex-M33 companion core, with \qty{2}{\mega\byte} of \ac{SRAM} and no \ac{NPU} (the \acs{NPU}-equipped variant is the RA8P1).

\textit{\textbf{Sony IMX500}}~\citep{imx500}.
The IMX500 is a stacked vision sensor that integrates a dedicated \ac{AI} processing unit on the same die as the \ac{CMOS} image array, so preprocessing and inference run on the sensor and the image never has to leave it.
Its processor runs at \qty{262.5}{\mega\hertz}.
Previous work has exploited this in-sensor digital architecture for gaze estimation~\citep{Bonazzi2023}, promptable image segmentation~\citep{Bonazzi2025, Bonazzi2026a}, industrial anomaly detection~\citep{Bonazzi2026b}, agricultural disease detection~\citep{Tong2026}, Earth Observation~\citep{Capogrosso2026a}, and recognition of retail products at the edge~\citep{Bommi2026}.

\textit{\textbf{STMicroelectronics STM32N6}}~\citep{Desoli2026}.
Released in 2025, the STM32N6 is STMicroelectronics' first \ac{MCU} to integrate an in-house Neural-ART \ac{NPU}, clocked at \qty{1}{\giga\hertz}, alongside an \qty{800}{\mega\hertz} Cortex-M55 core with Helium vector extensions.
Because the device is new, four works have used it so far: Bartoli~\textit{et al.}~\citep{Bartoli2025} propose a framework for benchmarking \ac{TinyML} latency and energy on it, Capogrosso~\textit{et al.}~\citep{Capogrosso2026b} use it for space-related applications, Giordano~\textit{et al.}~\citep{Giordano2025} demonstrate sub-millisecond event-based eye tracking and Manzo~\textit{et al.}~\citep{manzo2026} exploit it for hand gesture recognition.

\textit{\textbf{STMicroelectronics STM32U5}}~\citep{stm32u5}.
The STM32U5 is an ultra-low-power Cortex-M33 device with no dedicated neural accelerator and no Helium vector extension: its DSP extension offers only 32-bit \ac{SIMD} instructions (four 8-bit or two 16-bit operations each), and inference runs entirely on the \ac{CPU} at up to \qty{160}{\mega\hertz} with \qty{3}{\mega\byte} of \ac{SRAM} (STM32U5F7).
The sole published deployment we found reports tiny neural networks for depth-shape detection of mice in constrained cage environments~\citep{Pau2024}.

\textit{\textbf{STMicroelectronics STM32V8}}~\citep{stm32v8}.
The STM32V8 is built around an \qty{800}{\mega\hertz} Arm Cortex-M85 core with Helium vector extensions and \qty{1.5}{\mega\byte} of on-chip \ac{SRAM}.
It is the first \ac{MCU} manufactured in an \qty{18}{\nano\metre} FD-SOI process, with up to \qty{4}{\mega\byte} of embedded phase-change memory for code and weights.
Like the RA8M2 it has no dedicated \ac{NPU}: inference relies on the Helium datapath of the \ac{CPU}, which executes 8-bit multiply-accumulate operations in \ac{SIMD} fashion.
It draws the highest measured power in our study (\qty{343.5}{\milli\watt} average during inference, Fig.~\ref{fig:power}), but its high clock rate keeps latency at \qtyrange{15.6}{22.2}{\milli\second} for the three models it can run.
EfficientNet could not be deployed on it: its \qty{3.97}{\mega\byte} of INT8 weights, together with the runtime and application code, do not fit in the \qty{4}{\mega\byte} of on-chip non-volatile memory (Table~\ref{tab:experimental}).

\subsection{Vision Models} \label{subsec:models}

We choose the models to vary the three properties that decide how a network runs on edge hardware: how much it computes (MACs), how much memory it needs (weights and activations), and how many MACs it performs per byte moved (arithmetic intensity).

\textit{\textbf{Models.}}
Four \ac{CNN} image classifiers form the core benchmark and run on the nine devices other than the Alif E8, wherever memory and toolchain allow: \textit{i) MobileNetV3}~\citep{Howard2019}, a \ac{NAS}-designed depthwise-separable network for mobile processors; \textit{ii) MCUNetV1}~\citep{Lin2020}, co-designed with its runtime for sub-\qty{1}{\mega\byte} \ac{SRAM}; \textit{iii) EfficientNet}~\citep{Tan2019}, a compound-scaled network and the largest of our \acp{CNN}, which runs on seven of the nine devices that run the \acp{CNN}; \textit{iv) SqueezeNet}~\citep{Iandola2016}, a fire-module network with the most MACs of the four \acp{CNN}.
Two vision transformers are added on the devices built for attention workloads, the Alif E8 and Sony IMX500 (Section~\ref{subsec:latency}): \textit{i) DeiT-tiny}~\citep{Touvron2021}, a pure transformer of twelve attention blocks over 197 tokens, and \textit{ii) MobileViT-XXS}~\citep{Mehta2022}, a hybrid that interleaves convolutional blocks with transformer layers.

\textit{\textbf{Deployment.}}
The \acp{CNN} take a $64\times64$ RGB input and are trained on the ten-class EuroSAT land-cover dataset~\citep{Helber2019}; the MCUNetV1 checkpoint keeps its 1000-class classifier (\qty{2}{\percent} of its MACs).
The transformers run at $224\times224$ (DeiT-tiny) and $192\times192$ (MobileViT-XXS) with 1000-class classifiers.

\begin{table*}[t]
  \centering
  \caption{Structural properties of the benchmark models, sorted by intensity
           $I$ within each family ($\downarrow$: increasing downward).
           Squares: model colours of the figures.}
  \label{tab:models}
  \small
  \setlength{\tabcolsep}{4pt}
  \newsavebox{\modeltablebox}
  \sbox{\modeltablebox}{%
  \begin{tabular}{@{}l c c rr rrr rr rrrrr r@{}}
    \toprule
    & & & \multicolumn{5}{c}{Memory}
      & \multicolumn{8}{c}{Compute} \\
    \cmidrule(lr){4-8} \cmidrule(l){9-16}
    & & & Params & Peak act. & \multicolumn{3}{c}{Traffic $Q$ (MB)} & & Conv & \multicolumn{5}{c}{MACs by layer type (\%)} & Intensity $I$\,$\downarrow$ \\
    \cmidrule(lr){6-8} \cmidrule(lr){11-15}
    Model & Input & Output
      & (M) & (kB) & Weights & Act. & Total
      & MMAC & layers
      & $k{\times}k$ conv & $1{\times}1$ conv & Depthwise & FC & Attention
      & (MAC/B) \\
    \midrule
    \multicolumn{16}{@{}l}{\textit{CNNs}} \\
    \devsq{mobilenetcolor}MobileNetV3~\citep{Howard2019} & $64{\times}64$ & $10{\times}1$ & 1.52 & 36 & 1.51 & 0.25 & 1.76 & 6.1 & 52 & 8 & 70 & 11 & 11 & -- & 3.5  \\
    \rowrule
    \devsq{efficientnetcolor}EfficientNet~\citep{Tan2019} & $64{\times}64$ & $10{\times}1$ & 4.00 & 288 & 3.97 & 1.13 & 5.10 & 32.2 & 81 & 3 & 88 & 9 & $<$1 & -- & 6.3  \\
    \rowrule
    \devsq{mcunetcolor}MCUNetV1~\citep{Lin2020} & $64{\times}64$ & $1000{\times}1$ & 1.72 & 96 & 1.71 & 0.91 & 2.62 & 20.4 & 51 & 4 & 76 & 18 & 2 & -- & 7.8  \\
    \rowrule
    \devsq{squeezenetcolor}SqueezeNet~\citep{Iandola2016} & $64{\times}64$ & $10{\times}1$ & 0.74 & 158 & 0.74 & 0.48 & 1.22 & 51.6 & 26 & 83 & 17 & 0 & 0 & -- & 42.3 \\
    \midrule
    \multicolumn{16}{@{}l}{\textit{Vision transformers}} \\
    MobileViT-XXS~\citep{Mehta2022} & $192{\times}192$ & $1000{\times}1$ & 1.27 & 1\,009 & 1.26 & 8.35 & 9.61 & 210.0 & 35 & 30 & 20 & 3 & 36 & 12 & 21.9 \\
    \rowrule
    DeiT-tiny~\citep{Touvron2021} & $224{\times}224$ & $1000{\times}1$ & 5.72 & 480 & 5.65 & 12.06 & 17.71 & 1\,253.7 & 1 & 2 & 0 & 0 & 83 & 14 & 70.8 \\
    \bottomrule
  \end{tabular}}
  \ifdim\wd\modeltablebox>\textwidth
    \resizebox{\textwidth}{!}{\usebox{\modeltablebox}}%
  \else
    \usebox{\modeltablebox}%
  \fi
\end{table*}

\textit{\textbf{Structural profile.}}
Table~\ref{tab:models} characterizes the models from their exported ONNX graphs, independently of any toolchain.
It splits the MACs by the type of layer that performs them: $k\times k$ convolutions with $k>1$, $1\times1$ (pointwise) and depthwise convolutions, fully connected (FC) layers, and the attention products $QK^\top$ and $AV$.
It also estimates peak activation memory as the largest sum of simultaneously live INT8 tensors.
Finally, it gives the total memory traffic $Q$, the minimum number of bytes a model moves during one inference.
For every convolution, fully connected layer and attention product, $Q$ counts the INT8 bytes of the layer's input activations, its weights and its output activation, each once.
$Q$ is therefore the sum of two terms, both listed in Table~\ref{tab:models}: a weight term, equal to the INT8 weight storage of these layers, and an activation term, in which a tensor written by one layer is counted again when the next layer reads it.
Because it assumes that no weight or activation is fetched twice, $Q$ is a lower bound on the traffic a device generates.
The intensity $I=\text{MAC}/Q$, the operational intensity of the Roofline model~\citep{Williams2009}, is the number of MACs a model performs per byte of total memory traffic.

Four properties of these models matter for the results.
\begin{enumerate}
  \item \textit{MACs and parameters do not track each other.} SqueezeNet has the fewest parameters (0.74\,M) but the most \ac{CNN} MACs, because its weights act at high resolution, whereas EfficientNet has the most parameters (4.00\,M) but applies its widest layers to $2\times2$ and $4\times4$ maps.
  Weight storage spans \qtyrange{0.74}{4.0}{\mega\byte} and peak activation memory \qtyrange{36}{288}{\kilo\byte} across the \acp{CNN}.
  \item \textit{Where the MACs are spent differs by family.} The three MBConv networks (MobileNetV3, MCUNetV1 and EfficientNet, built from inverted-residual blocks) spend \qtyrange{70}{88}{\percent} of their MACs in pointwise and \qtyrange{9}{18}{\percent} in depthwise convolutions, spread evenly over depth; SqueezeNet spends \qty{83}{\percent} in standard $k\times k$ convolutions, concentrated in its stem and two fire modules; the transformers spend \qty{83}{\percent} (DeiT-tiny) and \qty{36}{\percent} (MobileViT-XXS) in FC layers and \qtyrange{12}{14}{\percent} in attention.
  \item \textit{Weights dominate the traffic of the \acp{CNN}.} At $64\times64$ input, weights make up \qtyrange{61}{86}{\percent} of $Q$ for the \acp{CNN}, whereas activations make up \qty{68}{\percent} (DeiT-tiny) and \qty{87}{\percent} (MobileViT-XXS) for the transformers, which run at higher resolution over 197 or more tokens.
  \item \textit{Intensity spans $12\times$.} The MBConv networks perform 3.5--7.8 MACs per byte of total memory traffic and SqueezeNet 42.3, so the models cover both compute-heavy and data-heavy models (Section~\ref{sec:discussion}).
\end{enumerate}

\subsection{Methodology} \label{subsec:methodology}

\subsubsection{Deployment Protocol}

Devices run at the clock frequencies in Table~\ref{tab:hardware}; on \acs{NPU}-equipped devices the model runs on the \ac{NPU} and any operator the compiler cannot map falls back to the \ac{CPU}.
The per-device flows are as follows.

\textit{\textbf{Alif E7 and E8.}}
Models are converted to fully integer INT8 TFLite and compiled with Arm's Vela compiler, which replaces every supported operator with an \ac{NPU} command stream and leaves the rest to the Cortex-M55.
On the E7 we target the high-performance Cortex-M55 and its 256-MAC Ethos-U55 (compiled by Vela with Arm's high-end embedded system model and the shared-\ac{SRAM} memory mode, in which the \ac{CPU} and \ac{NPU} share the activation \ac{SRAM}); on the E8 we target the 256-MAC Ethos-U85.
The compiled model is linked into a TFLite Micro application built with Alif's ML embedded evaluation kit, with weights stored in MRAM and the activation arena in on-chip \ac{SRAM} sized from Vela's memory report; the inference call is timed with the Cortex-M55 DWT cycle counter.

\textit{\textbf{Ambiq Apollo510B.}}
Models are converted to INT8 TFLite and deployed with Ambiq's neuralSPOT SDK, which runs them under TFLite Micro on the Cortex-M55 of the Apollo510B evaluation kit at \qty{250}{\mega\hertz}.

\textit{\textbf{GreenWaves GAP9.}}
Models are deployed with the GAP SDK.
NNTool imports the network, applies post-training INT8 quantization from calibration statistics and fuses operators; the Autotiler then generates C code that tiles each layer so that weights and activations are moved between the memory levels (L3, L2 and the cluster L1) by \ac{DMA} while the eight-core cluster and the NE16 compute.

\textit{\textbf{Renesas RA8M2.}}
Models are converted to INT8 TFLite and deployed with Renesas e\textsuperscript{2}~studio and the Flexible Software Package, which provides TFLite Micro with CMSIS-NN kernels, so that the network runs bare metal on the Helium datapath of the Cortex-M85 at \qty{1}{\giga\hertz}; the inference call is timed with the DWT cycle counter.

\textit{\textbf{Sony IMX500.}}
Models are quantized to INT8 with Sony's Model Compression Toolkit, converted with the IMX500 converter and packaged into a firmware file that is loaded into the sensor; inference then runs on every captured frame, without involvement of the host processor.

\textit{\textbf{STM32N6.}}
The ONNX models are quantized with the ONNX Runtime static pipeline (QDQ format, signed INT8 weights and activations, per-channel weights), with no manual graph rewriting beyond fixing the batch dimension to one and replacing two dynamic-shape patterns with static equivalents.
They are compiled with ST Edge AI Core v4.0.0, with the device in Overdrive mode (\qty{800}{\mega\hertz} \ac{CPU}, \qty{1}{\giga\hertz} \ac{NPU}).

\textit{\textbf{STM32U5.}}
Models are converted to INT8 TFLite and compiled into C code with ST's X-CUBE-AI, and the generated network runs on the Cortex-M33 at \qty{160}{\mega\hertz}; the inference call is timed with the DWT cycle counter.

\textit{\textbf{STM32V8.}}
Models run on the Cortex-M85 at \qty{800}{\mega\hertz} under TFLite Micro; each output is checked against a reference output computed off-device.

\subsubsection{Metrics}

For every model and device, we report:
\begin{itemize}
  \item \textit{Latency} $t_{\text{inf}}$ (ms): the time of one forward pass.
  \item \textit{Energy} $E=\bar{P}\,t_{\text{inf}}$ (mJ), with $\bar{P}$ the mean power during inference, and the \textit{energy above idle} $\Delta E=(\bar{P}-P_{\text{idle}})\,t_{\text{inf}}$, which removes the static power of everything the measurement includes besides the compute and so compares devices measured on different supply domains.
  \item \textit{Compute efficiency}, $\text{MAC/cycle}=\text{MACs}/(f_{\text{clk}}\,t_{\text{inf}})$, with $f_{\text{clk}}$ the clock of the unit that runs the model (the \ac{NPU} clock on \acs{NPU}-equipped devices): how much work a device completes per cycle, independently of its clock rate.
  \item \textit{Traffic rate}, $\text{B/cycle}=Q/(f_{\text{clk}}\,t_{\text{inf}})$, with $Q$ the total memory traffic of Table~\ref{tab:models}: the bytes of weights and activations moved per cycle of the same clock. Because $Q$ counts each tensor once, it is a lower bound on the bytes the hardware moves, summed over the memories that hold weights and activations; it is nearly constant across models on a memory-bound device (Section~\ref{sec:discussion}).
  \item \textit{Energy efficiency}, $\text{MMAC/J}=\text{MMAC}/\Delta E$: work per joule above idle, comparable across models of different size.
\end{itemize}

\textit{\textbf{Scaling analysis.}}
To test whether a device is limited by compute or by data movement, we fit, per device that runs the \acp{CNN} (all but the Alif E8) and over the $n$ \acp{CNN} it runs ($n=3$ or 4), two power laws, $t_{\text{inf}}\propto\text{MAC}^{\beta_{\text{MAC}}}$ and $t_{\text{inf}}\propto Q^{\beta_Q}$, by ordinary least squares in log space, $\log t_{\text{inf},i}=a+\beta\log x_i$, with $x$ the model's MAC count or its total memory traffic $Q$ (Table~\ref{tab:models}) and $t_{\text{inf}}$ the latency of Table~\ref{tab:experimental} ($t_{\text{DSP}}$ on the IMX500).
We report each exponent with $r$, the Pearson correlation of $\log x$ and $\log t_{\text{inf}}$.
We fit the two separately because MAC count and $Q$ are uncorrelated across the four \acp{CNN} ($r=0.08$); with three or four points per device, $\beta$ and $r$ describe the trend and are not significance tests.

\subsubsection{Latency}

On eight of the ten devices, a cycle counter or hardware timer brackets the inference call (the DWT cycle counter on the Cortex-M devices), and latency is averaged over ten consecutive inferences (100 on the GAP9, at least 160 on the Alif E7) after a warm-up run.
Two devices differ: on the STM32N6 a GPIO pin is raised for the duration of each inference, and the pulse length is averaged over ten inferences; and on the Sony IMX500, where inference runs inside the sensor, latency is read from the sensor's own counters.

\textit{\textbf{Sony IMX500.}}
The sensor reports two inference-time counters with every frame.
We use $t_{\text{DSP}}$, the DSP runtime counter, taken as the median over 120 frames after 30 warm-up frames; it is stable to a few $\mu$s for each network.
The second counter resolves only whole milliseconds: per frame it equals $t_{\text{DSP}}$ rounded up to the next millisecond plus a \qtyrange{0}{1}{\milli\second} phase that cycles with the frame period, and its whole-millisecond part equals $\lceil t_{\text{DSP}}\rceil$ for 28 of the 29 networks we ran.
Both are on-sensor times and exclude the transfer of the output tensor to the host.

\subsubsection{Power}

We measure power on six devices; for the other four we report latency and MAC/cycle only.

\begin{figure*}[!t]
  \centering
  \includegraphics[width=\textwidth]{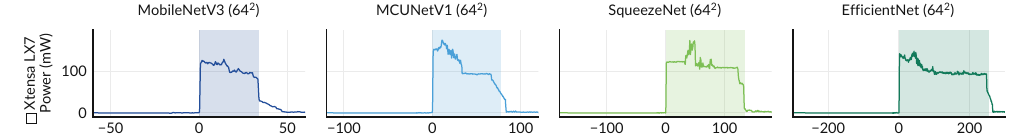}\\[0pt]
  \includegraphics[width=\textwidth]{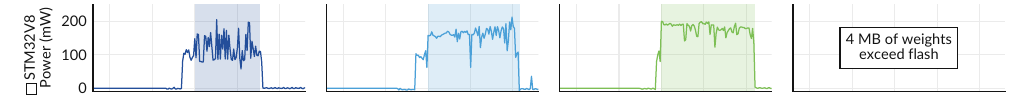}\\[0pt]
  \includegraphics[width=\textwidth]{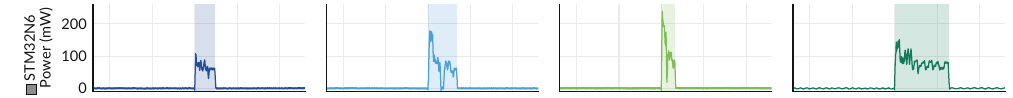}\\[0pt]
  \includegraphics[width=\textwidth]{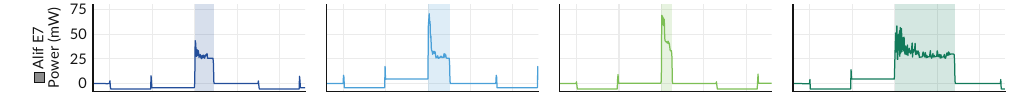}\\[0pt]
  \includegraphics[width=\textwidth]{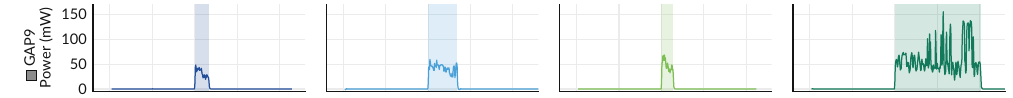}\\[0pt]
  \includegraphics[width=\textwidth]{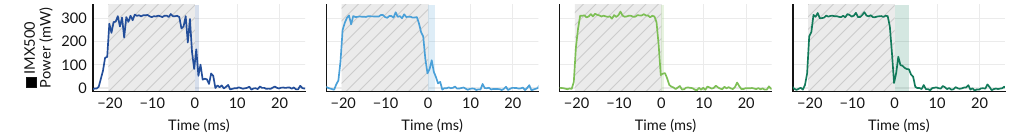}\\[0pt]
  \includegraphics[width=\textwidth]{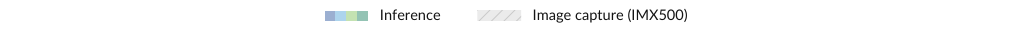}
  \caption{Power above idle per model, aligned at the start of inference (colored band): STM32N6 (VDDCORE rail),
           STM32V8 (supply; no EfficientNet, see Table~\ref{tab:experimental}), Alif E7 (J5 rail, mean of at least 160 runs), GAP9 (module supply, mean of 100 runs), IMX500 (camera module, mean of 901 frames)
           and Xtensa~LX7 (supply; own time axis per panel). Hatched grey: IMX500 image capture. The power axis differs per device.}
  \label{fig:power}
\end{figure*}

\textit{\textbf{STM32N6 and STM32V8.}}
On the STM32N6 a Nordic PPK2 samples the current of the VDDCORE rail (\qty{0.89}{\volt}) at \qty{100}{\kilo\hertz}, the GPIO pulse marks each inference, and $\bar{P}$ and $E$ are
averaged over ten inferences.
On the STM32V8 a Keysight N6705C DC power analyzer records the supply power at \qty{3.5}{\kilo\hertz}; with no marker, the inference window is the interval of length $t_{\text{inf}}$ with the highest integrated power.
Because the STM32N6 is measured on its core rail and the STM32V8 on its supply, we compare the two on $\Delta E$, with $P_{\text{idle}}$ the mean power \qtyrange{2}{6}{\milli\second} after the inference.
Fig.~\ref{fig:power} shows their traces next to those of the Alif E7, GAP9, IMX500 and Xtensa~LX7, for all four models on the STM32N6 and the three that fit on the STM32V8, aligned at the start of inference (STM32N6 drawn as a \qty{0.2}{\milli\second} moving average of its \qty{10}{\micro\second} samples).

\textit{\textbf{Sony IMX500.}}
The sensor and its processor share one die, so we measure the whole camera module on its 3V3\_SYS rail, using the power monitor of the host Raspberry~Pi~5 (about 25 samples/s).
This rate is too low to resolve a single frame, so we stream at \qty{5}{\fps} and average the samples of 901 frames by their position within the \qty{200}{\milli\second} frame period (Fig.~\ref{fig:power}; more networks in Fig.~S1 of the supplementary material).
Each frame starts with a burst of about \qty{320}{\milli\watt} while the image is read out, which lasts \qty{19.4}{\milli\second} (\qtyrange{19.1}{20.1}{\milli\second}) for every network, followed by inference.
We take the inference window as $t_{\text{DSP}}$ from where the inference power rises, \qty{1}{\milli\second} after the end of readout, and report the energy above idle, $\Delta E$, as for the STM32N6 and STM32V8, with $P_{\text{idle}}$ the power level outside the burst.

\textit{\textbf{Xtensa~LX7.}}
A Nordic PPK2 samples the supply current at \qty{100}{\kilo\hertz} over ten INT8 inferences, with the inference interval flagged by the firmware; as for the other devices, we report $E$ and $\Delta E$, with $P_{\text{idle}}$ the mean power before inference.

\textit{\textbf{GAP9.}}
A Nordic PPK2 samples the current of the GAP9 module supply (\qty{1.8}{\volt}) at \qty{100}{\kilo\hertz}, with the \ac{SoC} at \qty{370}{\mega\hertz} and \qty{0.8}{\volt}; a GPIO pin marks each inference, latency is read from the on-chip timer, and $\bar{P}$, $E$ and $\Delta E$ are averaged over 100 inferences after a warm-up run, with $P_{\text{idle}}$ the mean power \qtyrange{2}{6}{\milli\second} after the inference.

\textit{\textbf{Alif E7.}}
A Nordic PPK2 in ampere-meter mode samples the current of the \qty{3.3}{\volt} J5 rail of the DevKit-E7 at \qty{100}{\kilo\hertz}, with the Cortex-M55 and Ethos-U55 at \qty{400}{\mega\hertz}; a GPIO pin marks each inference, and $\bar{P}$, $E$ and $\Delta E$ are averaged over at least 160 inferences after a warm-up run, with $P_{\text{idle}}$ the mean power \qtyrange{2}{6}{\milli\second} after the inference.

\section{Experimental Results} \label{sec:results}

\begin{figure}[!t]
  \centering
  \includegraphics[width=\columnwidth]{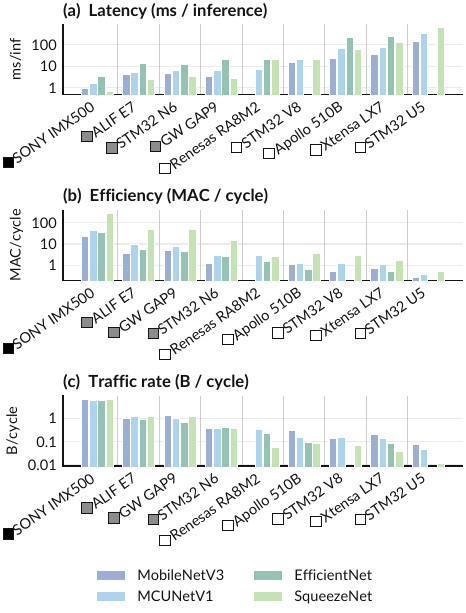}
  \caption{Latency, MAC/cycle and traffic rate B/cycle (log scale) of the four CNN models
           on the nine devices, sorted per panel by cross-model mean. Squares: device type and values as in Table~\ref{tab:experimental}.}
  \label{fig:histograms}
\end{figure}

\begin{figure}[!t]
  \centering
  \includegraphics[width=\columnwidth]{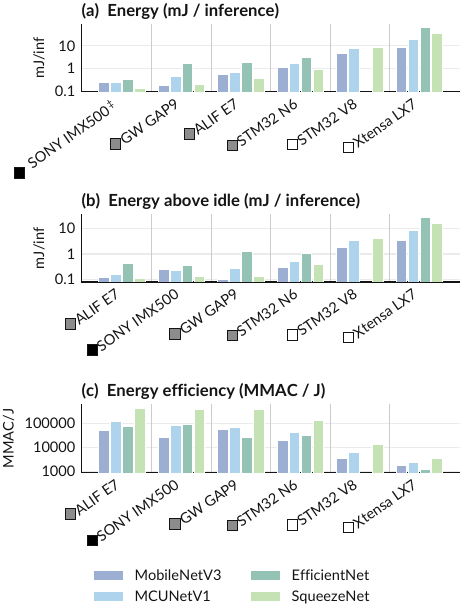}
  \caption{Energy per inference, energy above idle and MMAC/J on $\Delta E$ (log scale) on the six devices with
           measured energy; $^\ddagger$IMX500: energy above the standing level, as no total is reported. Values in Table~S2 of the supplementary material.}
  \label{fig:energy}
\end{figure}

Table~\ref{tab:experimental} reports latency, MAC/cycle and B/cycle, with devices sorted by ascending mean MAC/cycle and the two Alif devices listed together.
Fig.~\ref{fig:histograms} shows latency, MAC/cycle and B/cycle, and Fig.~\ref{fig:energy} shows energy per inference, energy above idle and MMAC/J.
Table~\ref{tab:bound} and Fig.~\ref{fig:bound} report the scaling analysis, discussed in Section~\ref{sec:discussion}.

\begin{table*}[t]
  \centering
  \caption{Latency, MAC/cycle and B/cycle per device and model; best in bold, second best underlined.
           ``---'': not deployable (STM32V8: EfficientNet does not fit in memory; RA8M2: hard-swish unsupported;
           STM32U5: unsupported operators). Empty: not evaluated (vision transformers only on the Alif E8 and IMX500).
           Device type: \cpusq MCU without NPU, \npusq MCU with NPU, \dspsq DSP.}
  \label{tab:experimental}
  \newcommand{\twoline}[2]{\begin{tabular}[b]{@{}c@{}}#1\\#2\end{tabular}}
  \resizebox{\textwidth}{!}{%
  \begin{tabular}{@{}l *{6}{rrr}@{}}
    \toprule
    & \multicolumn{3}{c}{\devsq{mobilenetcolor}\textbf{MobileNetV3}~\cite{Howard2019}}
      & \multicolumn{3}{c}{\devsq{mcunetcolor}\textbf{MCUNetV1}~\cite{Lin2020}}
      & \multicolumn{3}{c}{\devsq{efficientnetcolor}\textbf{EfficientNet}~\cite{Tan2019}}
      & \multicolumn{3}{c}{\devsq{squeezenetcolor}\textbf{SqueezeNet}~\cite{Iandola2016}}
      & \multicolumn{3}{c}{\textbf{DeiT-tiny}~\cite{Touvron2021}}
      & \multicolumn{3}{c}{\textbf{MobileViT-XXS}~\cite{Mehta2022}} \\
    & \multicolumn{3}{c}{64$\times$64} & \multicolumn{3}{c}{64$\times$64}
      & \multicolumn{3}{c}{64$\times$64} & \multicolumn{3}{c}{64$\times$64}
      & \multicolumn{3}{c}{224$\times$224} & \multicolumn{3}{c}{192$\times$192} \\
    \cmidrule(lr){2-4} \cmidrule(lr){5-7} \cmidrule(lr){8-10}
    \cmidrule(lr){11-13} \cmidrule(lr){14-16} \cmidrule(l){17-19}
    \textbf{Device}
      & ms/inf & \twoline{MAC/}{cycle} & \twoline{B/}{cycle} & ms/inf & \twoline{MAC/}{cycle} & \twoline{B/}{cycle} & ms/inf & \twoline{MAC/}{cycle} & \twoline{B/}{cycle}
      & ms/inf & \twoline{MAC/}{cycle} & \twoline{B/}{cycle} & ms/inf & \twoline{MAC/}{cycle} & \twoline{B/}{cycle} & ms/inf & \twoline{MAC/}{cycle} & \twoline{B/}{cycle} \\
    \midrule
    \cpusq STM32U5~\citep{stm32u5} & 140.7                     & 0.3                       & 0.08 & 357.3                     & 0.4                       & 0.05 & ---                        & ---                        & --- & 629.2                     & 0.5                        & 0.01 &                           &                             &  &                           & & \\
    \rowrule
    \cpusq Xtensa LX7~\citep{esp32s3} & 34.8                      & 0.7                       & 0.21 & 75.9                      & 1.1                       & 0.14 & 252.5                      & 0.5                       & 0.08 & 130.0                     & 1.7                        & 0.04 &                           &                             &  &                           & & \\
    \rowrule
    \cpusq STM32V8~\citep{stm32v8} & 15.6                      & 0.5                       & 0.14 & 21.5                      & 1.2                       & 0.15 & ---                        & ---                        & --- & 22.2                      & 2.9                        & 0.07 &                           &                             &  &                           & & \\
    \rowrule
    \cpusq Apollo510B~\citep{apollo510b} & 22.8                      & 1.1                       & 0.31 & 66.4                      & 1.2                       & 0.16 & 211.2                      & 0.6                       & 0.10 & 59.2                      & 3.5                        & 0.08 &                           &                             &  &                           & & \\
    \rowrule
    \cpusq RA8M2~\citep{ra8m2} & ---                       & ---                        & --- & 7.4                       & 2.8                       & 0.35 & 22.3                       & 1.5                       & 0.23 & 20.8                      & 2.5                        & 0.06 &                           &                             &  &                           & & \\
    \rowrule
    \npusq STM32N6~\citep{Desoli2026} & 4.8                       & 1.3                       & 0.37 & 6.8                       & 3.0                       & 0.38 & \underline{12.8} & 2.5                       & 0.40 & 3.3                       & 15.8                       & 0.37 &                           &                             &  &                           & & \\
    \rowrule
    \npusq GAP9~\citep{Flamand2018} & \underline{3.3} & \underline{5.0} & \underline{1.44} & 6.8 & 8.1 & 1.04 & 20.3 & 4.3 & 0.68 & 2.8 & 50.4 & 1.19 &                           &                             &  &                           & & \\
    \rowrule
    \npusq Alif E7~\citep{alife7} & 4.4 & 3.4 & 0.99 & \underline{5.1} & \underline{10.0} & \underline{1.28} & 14.2 & \underline{5.7} & \underline{0.90} & \underline{2.5} & \underline{51.0} & \underline{1.21} &                           &                             &  &                           & & \\
    \rowrule
    \npusq Alif E8~\citep{alife8} &                           &                            &  &                           &                            &  &                            &                            &  &                           &                             &  & \underline{174.0} & \underline{18.0} & \underline{0.25} & \underline{214.4} & \underline{2.5} & \underline{0.11} \\
    \rowrule
    \dspsq IMX500~\citep{imx500} & \textbf{1.0}              & \textbf{23.5}             & \textbf{6.77} & \textbf{1.7}              & \textbf{46.6}             & \textbf{5.99} & \textbf{3.3}               & \textbf{37.3}             & \textbf{5.92} & \textbf{0.7}              & \textbf{282.0}             & \textbf{6.65} & \textbf{15.7}            & \textbf{304.0}             & \textbf{4.30} & \textbf{12.7}            & \textbf{62.8} & \textbf{2.88} \\
    \bottomrule
  \end{tabular}%
  }
\end{table*}

\subsection{Latency} \label{subsec:latency}

The Sony IMX500 achieves the lowest latency on all four models (\qtyrange{0.7}{3.3}{\milli\second}), followed by the GAP9 on MobileNetV3 and the Alif E7 on MCUNetV1 and SqueezeNet.
In contrast, the STM32U5, a Cortex-M33 \acs{MCU}, exhibits the highest latency on every model it runs, consistent with a core that has neither a Helium vector extension nor an accelerator.
The STM32N6 runs all four models in \qtyrange{3.3}{12.8}{\milli\second}, on par with the other \acsp{MCU} with an \acs{NPU}, and is the second fastest on EfficientNet (\qty{12.8}{\milli\second}).
On the two vision transformers, the Alif E8 runs DeiT-tiny in \qty{174}{\milli\second} and MobileViT-XXS in \qty{214}{\milli\second}, and the IMX500 runs them in 15.7 and \qty{12.7}{\milli\second}, $11.1\times$ and $16.9\times$ faster.

\subsection{Energy Consumption}

The Sony IMX500 uses \qtyrange{0.13}{0.35}{\milli\joule} per inference, the GAP9 \qtyrange{0.18}{1.71}{\milli\joule} on its module supply and the Alif E7 \qtyrange{0.36}{1.79}{\milli\joule} on its J5 rail; the STM32N6 uses \qtyrange{0.93}{3.2}{\milli\joule} on its compute rail, 4--9$\times$ less than the STM32V8, whereas the STM32V8 and Xtensa~LX7 use \qtyrange{4.8}{64}{\milli\joule}.
On energy above idle, which discounts the different measurement domains, the IMX500 uses 1.3--3$\times$ less than the STM32N6 on every model (0.24/0.31, 0.24/0.50, 0.35/1.05 and 0.13/\qty{0.39}{\milli\joule} on MobileNetV3, MCUNetV1, EfficientNet and SqueezeNet); the GAP9 (\qtyrange{0.11}{1.23}{\milli\joule}) uses the least of all devices on MobileNetV3 and matches the IMX500 on SqueezeNet (\qty{0.13}{\milli\joule}), and the Alif E7 (\qtyrange{0.12}{0.45}{\milli\joule}) uses the least on MCUNetV1 and SqueezeNet (0.16 and \qty{0.12}{\milli\joule}).
However, the Alif E7 idles at \qty{95}{\milli\watt} on its rail, against about \qty{23}{\milli\watt} on the GAP9 module, so its total energy exceeds the GAP9's on every model.
The IMX500 uses 8--31$\times$, the Alif E7 15--34$\times$ and the STM32N6 6--10$\times$ less than the STM32V8 (\qtyrange{1.85}{3.98}{\milli\joule}, Fig.~\ref{fig:power}); the Xtensa~LX7 uses \qtyrange{3.6}{26}{\milli\joule} above idle.

The traces in Fig.~\ref{fig:power} show how this energy is spent.
Above idle, the Alif E7 (\qtyrange{28}{46}{\milli\watt}) and GAP9 (\qtyrange{32}{61}{\milli\watt}) draw the least during inference, against \qtyrange{65}{118}{\milli\watt} on the STM32N6, \qtyrange{104}{115}{\milli\watt} on the Xtensa~LX7, \qtyrange{107}{244}{\milli\watt} on the IMX500 and \qtyrange{120}{177}{\milli\watt} on the STM32V8.
On the Alif E7, STM32N6, STM32V8 and Xtensa~LX7, SqueezeNet draws the most, as the model with the highest MAC/cycle on these devices.
Power on the Alif E7 and STM32N6 peaks within about \qty{1}{\milli\second} of the start of inference and then settles, so that the first fifth of an inference draws 1.3--2.3$\times$ the power of its second half, whereas the STM32V8 stays within \qty{13}{\percent} on the same comparison.
The GAP9 differs on EfficientNet, where its peak comes mid-inference (\qty{155}{\milli\watt}, \qty{11}{\milli\second} in).
On the IMX500, image capture dominates the trace: at about \qty{320}{\milli\watt} for \qty{19.4}{\milli\second}, it costs about \qty{6}{\milli\joule}, 18--48$\times$ the energy of the inference above idle, and $\Delta E$ does not include it.

\subsection{MAC/cycle and B/cycle}

SqueezeNet achieves the highest MAC/cycle on every device that runs the \acp{CNN} except the Renesas~RA8M2 (Section~\ref{sec:discussion}).
The low-power \acsp{MCU} (STM32U5, Apollo510B) stay below \qty{2}{\mac\per\cycle} on every model except the Apollo510B on SqueezeNet (3.5), while the Sony IMX500 and GAP9 exceed \qty{4}{\mac\per\cycle} on every \ac{CNN}.
On the vision transformers, DeiT-tiny reaches \qty{18.0}{\mac\per\cycle} on the Alif E8 and 304.0 on the IMX500, the highest in our study, against 2.5 and 62.8 for MobileViT-XXS.
B/cycle (Fig.~\ref{fig:histograms}c), in contrast, varies little across models on the accelerators (\qtyrange{5.92}{6.77}{\byte\per\cycle} on the IMX500, \qtyrange{0.37}{0.40}{\byte\per\cycle} on the STM32N6), while their MAC/cycle varies by an order of magnitude (23.5--282 and 1.3--15.8 on the \acp{CNN}; Section~\ref{sec:discussion}).
The IMX500 has the highest traffic rate, followed by the GAP9 (\qtyrange{0.68}{1.44}{\byte\per\cycle}) and the Alif E7 (\qtyrange{0.90}{1.28}{\byte\per\cycle}); the \acsp{MCU} without an \acs{NPU} reach at most \qty{0.35}{\byte\per\cycle}, and their B/cycle varies up to $8\times$ across models (\qtyrange{0.01}{0.08}{\byte\per\cycle} on the STM32U5), because their latency is not set by data movement.

\subsection{Energy Efficiency (MMAC/J)}

On energy above idle, the leader depends on the model: the GAP9 on MobileNetV3 (\qty{57736}{\mmac\per\joule}), the Alif E7 on MCUNetV1 and SqueezeNet (124300 and \qty{444531}{\mmac\per\joule}) and the Sony IMX500 on EfficientNet (\qty{91522}{\mmac\per\joule}).
The IMX500 reaches \qtyrange{25352}{401010}{\mmac\per\joule}, the GAP9 \qtyrange{26133}{393969}{\mmac\per\joule}, the Alif E7 \qtyrange{49275}{444531}{\mmac\per\joule} and the STM32N6 \qtyrange{19478}{132232}{\mmac\per\joule}, whereas the STM32V8 and Xtensa~LX7 reach at most \qty{12981}{\mmac\per\joule}, the Xtensa~LX7 being the least efficient on all four models.
However, high MAC/cycle does not imply high MMAC/J: a device that draws high power can complete many MACs per cycle and still use more energy per inference than a slower, lower-power device.
Energy, like latency, follows the data moved: across the \acp{CNN}, the $\Delta E$ of the four accelerators grows with $Q$ ($r_Q=0.90$--0.95) and not with MACs ($|r_{\text{MAC}}|\le0.46$), whereas that of the STM32V8 and Xtensa~LX7 follows MACs ($r_{\text{MAC}}=0.96$ and 0.88, $r_Q=-0.15$ and 0.48).

\begin{figure}[!t]
  \centering
  \includegraphics[width=\columnwidth]{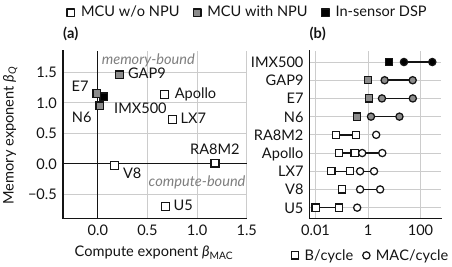}
  \caption{(a) Growth exponents of latency with MACs ($\beta_{\text{MAC}}$) and with memory traffic $Q$ ($\beta_Q$): accelerators lie near the memory bound $(0,1)$.
  (b) Range of B/cycle and MAC/cycle per device over the \acp{CNN}. Values in Table~\ref{tab:bound}.}
  \label{fig:bound}
\end{figure}

\begin{table}[!t]
  \centering
  \caption{Growth exponents $\beta$ and correlations $r$ of latency with MACs and with memory traffic $Q$
           ($t_{\text{inf}}\propto x^{\beta}$) over the $n$ \acp{CNN} each device runs, and the range of B/cycle and MAC/cycle.}
  \label{tab:bound}
  \resizebox{\columnwidth}{!}{%
  \begin{tabular}{@{}l c rr rr rr@{}}
    \toprule
    & & \multicolumn{2}{c}{\textbf{Compute}} & \multicolumn{2}{c}{\textbf{Memory}} & \\
    \cmidrule(lr){3-4} \cmidrule(lr){5-6}
    \textbf{Device} & $n$ & $\beta_{\text{MAC}}$ & $r_{\text{MAC}}$
      & $\beta_Q$ & $r_Q$ & B/cycle & MAC/cycle \\
    \midrule
    \multicolumn{8}{@{}l}{\textit{MCU without NPU}} \\
    \cpusq STM32U5~\citep{stm32u5} & 3 & 0.68 & 1.00 & $-0.70$ & $-0.35$ & 0.01--0.08 & 0.3--0.5 \\
    \rowrule
    \cpusq Xtensa LX7~\citep{esp32s3} & 4 & 0.75 & 0.85 & 0.73 & 0.53 & 0.04--0.21 & 0.5--1.7 \\
    \rowrule
    \cpusq STM32V8~\citep{stm32v8} & 3 & 0.17 & 0.95 & $-0.03$ & $-0.06$ & 0.07--0.15 & 0.5--2.9 \\
    \rowrule
    \cpusq Apollo510B~\citep{apollo510b} & 4 & 0.67 & 0.70 & 1.14 & 0.76 & 0.08--0.31 & 0.6--3.5 \\
    \rowrule
    \cpusq RA8M2~\citep{ra8m2} & 3 & 1.18 & 0.84 & 0.01 & 0.01 & 0.06--0.35 & 1.5--2.8 \\
    \midrule
    \multicolumn{8}{@{}l}{\textit{MCU with NPU}} \\
    \npusq STM32N6~\citep{Desoli2026} & 4 & 0.02 & 0.03 & 0.95 & 1.00 & 0.37--0.40 & 1.3--15.8 \\
    \rowrule
    \npusq Alif E7~\citep{alife7} & 4 & $-0.01$ & $-0.01$ & 1.15 & 0.98 & 0.90--1.28 & 3.4--51.0 \\
    \rowrule
    \npusq GAP9~\citep{Flamand2018} & 4 & 0.22 & 0.23 & 1.46 & 0.99 & 0.68--1.44 & 4.3--50.4 \\
    \midrule
    \multicolumn{8}{@{}l}{\textit{In-sensor DSP}} \\
    \dspsq IMX500~\citep{imx500} & 4 & 0.06 & 0.09 & 1.10 & 1.00 & 5.92--6.77 & 23.5--282 \\
    \bottomrule
  \end{tabular}%
  }
\end{table}

\section{Discussion} \label{sec:discussion}

A device is limited either by how fast it computes or by how fast it moves data~\citep{Williams2009}; in Fig.~\ref{fig:bound}, a compute-bound device sits near $(\beta_{\text{MAC}},\beta_Q)=(1,0)$ and a memory-bound one near $(0,1)$.
All four accelerators sit near the memory bound: on the IMX500, GAP9, Alif E7 and STM32N6, latency follows the memory traffic $Q$, not the MAC count (Table~\ref{tab:bound}).
On the IMX500 and STM32N6, B/cycle changes by at most \qty{14}{\percent} while the MAC count changes by $8.5\times$, so these accelerators spend most of their time waiting for data, and at $64\times64$ input most of these data are weights (Table~\ref{tab:models}).
We establish this for TinyML-scale \acp{CNN} at $64\times64$ input, the regime these sub-watt devices target; larger models or higher input resolutions remain to be tested.
The \acsp{MCU} without an \acs{NPU} behave differently.
The STM32U5 and RA8M2 follow the MAC count, latency on the Xtensa~LX7 and Apollo510B grows with both quantities, and on the STM32V8 the MAC/cycle grows with model size (Table~\ref{tab:bound}).

Because the accelerators wait for data, their MAC/cycle is highest on the model of highest intensity: SqueezeNet (\qty{42.3}{\mac\per\byte}) reaches 5.1--6.2$\times$ the MAC/cycle of MCUNetV1 on these devices (Table~\ref{tab:experimental}).
As a consequence, MAC count, by which hardware-aware design ranks networks~\citep{Howard2019, Lin2020}, can rank them in the wrong order: on the Alif E7, SqueezeNet has $8.5\times$ the MACs of MobileNetV3 and finishes $1.8\times$ faster.
The vision transformers show the same effect: DeiT-tiny has $6\times$ the MACs of MobileViT-XXS but $3\times$ its intensity, and reaches 5--7$\times$ its MAC/cycle.
For \ac{NPU} targets, \ac{NAS} and compression should therefore minimize memory rather than MACs, and favor layers that reuse their weights, such as standard convolutions, over long chains of pointwise and depthwise layers.

The same reasoning tells when an \ac{NPU} is worth its cost.
Its advantage is largest on high-intensity models: the Alif E7 is only 1.4--1.6$\times$ faster than the \qty{1}{\giga\hertz} RA8M2 on the MBConv networks, but $8.2\times$ faster on SqueezeNet (Table~\ref{tab:experimental}).
It is also needed under tight latency or energy budgets, since no \acs{MCU} without an \acs{NPU} runs any model in less than \qty{7}{\milli\second}, and among the devices with measured energy the IMX500 has the lowest latency on every model, while the lowest energy above idle falls to the GAP9 on MobileNetV3, the Alif E7 on MCUNetV1 and SqueezeNet, and the IMX500 on EfficientNet.
Without such budgets, a fast vector \ac{CPU} can suffice for low-intensity MBConv networks: the \qty{1}{\giga\hertz} Cortex-M85 of the RA8M2 runs MCUNetV1 in \qty{7.4}{\milli\second}, against \qty{5.1}{\milli\second} on the Alif E7 (Table~\ref{tab:experimental}).

\section{Conclusion} \label{sec:conclusion}

We measured ten commercial edge devices, from an in-sensor \ac{DSP} and \ac{MCU}-integrated \acsp{NPU} to a RISC-V and high-clock Cortex-M85 \acsp{CPU}, on four \acp{CNN} and two vision transformers, each deployed with its vendor's toolchain, including four novel devices.
The central finding is that every sub-watt accelerator we measured on the \acp{CNN} is memory-bound: its latency follows the bytes of weights and activations of the model, not its MACs, whereas no \acs{MCU} without an \acs{NPU} behaves this way.
It follows that the \ac{NPU} advantage depends on the model's intensity, from only $1.4\times$ over a \qty{1}{\giga\hertz} Cortex-M85 on MCUNetV1 to $8.2\times$ on the high-intensity SqueezeNet, and that MAC count is a misleading cost proxy for these accelerators: models designed for them should minimize memory traffic.
The accelerators also pay off in energy: the Sony IMX500, GAP9, Alif E7 and STM32N6 use 6--34$\times$ less energy above idle than the STM32V8, and the Alif E7 and IMX500 form the power--performance Pareto front.

\section*{Acknowledgments}

This work was funded by the Swiss National Science Foundation (Grant 219943).

\bibliographystyle{paperrefs}
\bibliography{main}

\section*{AI Usage Disclosure}

We used Claude Opus 5.5, a large language model accessed through its chat interface, for two purposes during the preparation of this submission.

After the authors wrote the complete manuscript, we gave the model its \LaTeX{} source and asked it to list the presentation issues.
The model returned a list of suggested edits, each given as the current text and a proposed replacement.
The suggestions covered consistent device naming (e.g., ``STM32N6'' instead of ``STM32~N6''), grammar and punctuation, consistent spelling, acronym usage, figure-reference style, and the internal consistency of counts and claims across sections.
The authors reviewed each suggestion and applied, modified, or rejected it by hand; the model never edited the manuscript directly.

We also asked the model for general comments on the manuscript's clarity; the authors decided on and made all resulting revisions.
No generative AI tools were used for ideation, experiment design, deployment, measurement, data analysis, figure generation, or literature search.
The authors wrote all text, ran all experiments, selected and verified all references, and take full responsibility for the content of this paper.

\twocolumn[
\mlsystitle{Supplementary Material\\TinyML Hits the Memory Wall: A Benchmark of Ten Sub-Watt Edge Devices} 
\vskip 0.3in
]

\begin{figure*}[!h]
  \centering
  \includegraphics[width=\textwidth]{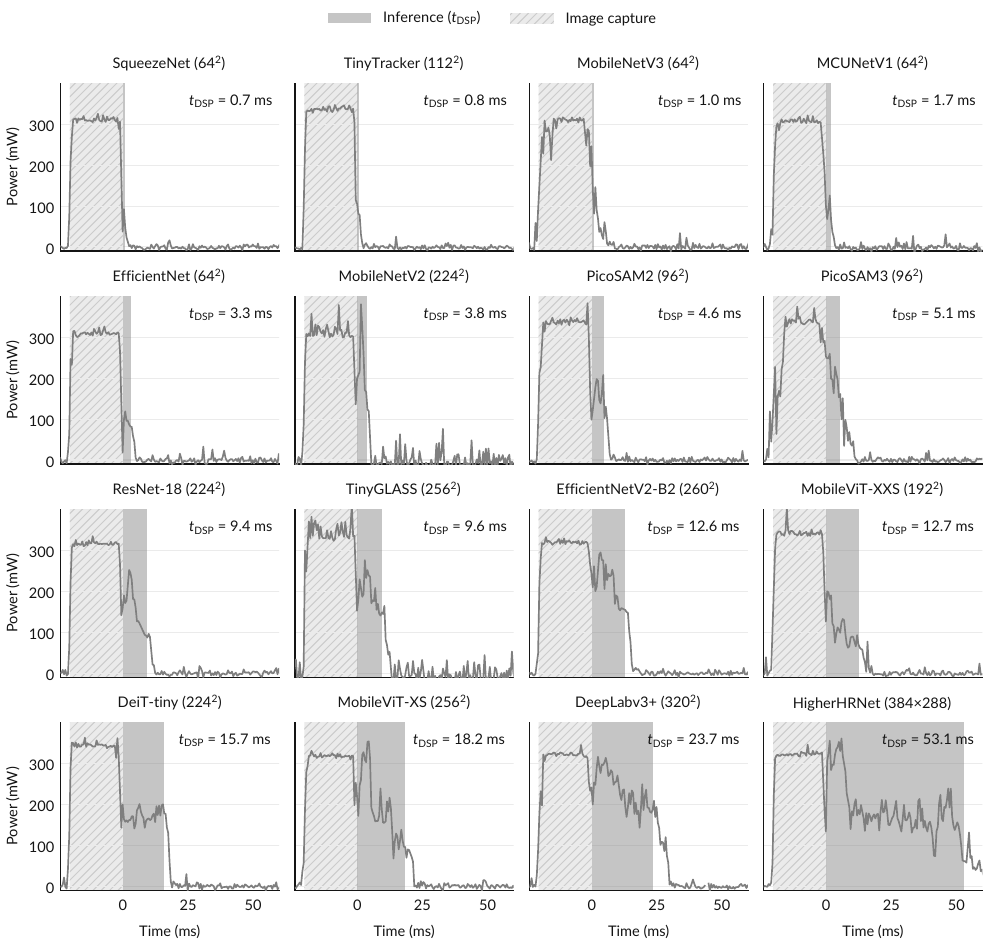}
  \caption{Sony IMX500 camera-module power above idle within one frame (mean of 901 frames in \qty{0.5}{\milli\second} bins)
           for our four models and 12 further networks: the two vision transformers, four in-sensor networks from prior work and six networks of the IMX500 model zoo, sorted by $t_{\text{DSP}}$.
           Hatched grey band: image capture (sensor readout, \qty{19.4}{\milli\second}, and the \qty{1}{\milli\second} until inference starts);
           shaded band: inference lasting $t_{\text{DSP}}$.}
  \label{fig:imx500_ladder}
\end{figure*}

Fig.~\ref{fig:imx500_ladder} extends the IMX500 row of Fig.~2 of the main paper to 12 further networks, sorted by $t_{\text{DSP}}$: the two vision transformers (DeiT-tiny and MobileViT-XXS), four in-sensor networks from prior work, TinyTracker~\citep{Bonazzi2023} for gaze estimation, PicoSAM2~\citep{Bonazzi2025} and PicoSAM3~\citep{Bonazzi2026a} for segmentation, and TinyGLASS~\citep{Bonazzi2026b} for anomaly detection, and six networks of the IMX500 model zoo~\citep{imx500models}.
The image capture lasts the same time for every network, while the inference power after it grows with $t_{\text{DSP}}$.
Table~\ref{tab:imx500_supp} lists the latency $t_{\text{DSP}}$ and the energy above idle $\Delta E$ per inference of all 16 networks, with their parameters and MACs.

\begin{table}[t]
  \centering
  \caption{The 16 networks of Fig.~\ref{fig:imx500_ladder} on the Sony IMX500, grouped by task and sorted by $t_{\text{DSP}}$ within each task:
           parameters, MACs, latency $t_{\text{DSP}}$ and energy above idle $\Delta E$ per inference ($\pm$ standard error).
           Source: this work, the venue and year of the prior in-sensor work, or the IMX500 model zoo (Sony). ``--'': unknown, as the model zoo does not pin the variant.}
  \label{tab:imx500_supp}
  \footnotesize
  \setlength{\tabcolsep}{3pt}
  \newsavebox{\imxtablebox}
  \sbox{\imxtablebox}{%
  \begin{tabular}{@{}l l c rr r r@{}}
    \toprule
    & & & Params & & $t_{\text{DSP}}$ & $\Delta E$ \\
    Network & Source & Input & (M) & MMAC & (ms/inf) & (mJ/inf) \\
    \midrule
    \multicolumn{7}{@{}l}{\textit{Classification}} \\
    \devsq{squeezenetcolor}SqueezeNet~\citep{Iandola2016} & This work & $64{\times}64$ & 0.74 & 51.6 & 0.70 & 0.13 $\pm$ 0.02 \\
    \devsq{mobilenetcolor}MobileNetV3~\citep{Howard2019} & This work & $64{\times}64$ & 1.52 & 6.1 & 0.99 & 0.24 $\pm$ 0.02 \\
    \devsq{mcunetcolor}MCUNetV1~\citep{Lin2020} & This work & $64{\times}64$ & 1.72 & 20.4 & 1.67 & 0.24 $\pm$ 0.03 \\
    \devsq{efficientnetcolor}EfficientNet~\citep{Tan2019} & This work & $64{\times}64$ & 4.00 & 32.2 & 3.28 & 0.35 $\pm$ 0.04 \\
    MobileNetV2 & Sony~\citep{imx500models} & $224{\times}224$ & 3.50 & 300.8 & 3.83 & 0.88 $\pm$ 0.06 \\
    ResNet-18 & Sony~\citep{imx500models} & $224{\times}224$ & 11.69 & 1\,814.1 & 9.40 & 1.49 $\pm$ 0.05 \\
    EfficientNetV2-B2 & Sony~\citep{imx500models} & $260{\times}260$ & 10.10 & 1\,697.5 & 12.61 & 2.84 $\pm$ 0.06 \\
    MobileViT-XXS~\citep{Mehta2022} & This work & $192{\times}192$ & 1.27 & 210.0 & 12.74 & 1.49 $\pm$ 0.06 \\
    DeiT-tiny~\citep{Touvron2021} & This work & $224{\times}224$ & 5.72 & 1\,253.7 & 15.71 & 2.65 $\pm$ 0.04 \\
    MobileViT-XS & Sony~\citep{imx500models} & $256{\times}256$ & 2.32 & 1\,025.9 & 18.20 & 3.77 $\pm$ 0.10 \\
    \midrule
    \multicolumn{7}{@{}l}{\textit{Segmentation}} \\
    PicoSAM2~\citep{Bonazzi2025} & Sensors'25 & $96{\times}96$ & 1.26 & 323.1 & 4.60 & 0.72 $\pm$ 0.04 \\
    PicoSAM3~\citep{Bonazzi2026a} & JSEN'26 & $96{\times}96$ & 1.37 & 330.2 & 5.12 & 1.18 $\pm$ 0.05 \\
    DeepLabv3+ & Sony~\citep{imx500models} & $320{\times}320$ & -- & -- & 23.66 & 5.73 $\pm$ 0.08 \\
    \midrule
    \multicolumn{7}{@{}l}{\textit{Gaze estimation}} \\
    TinyTracker~\citep{Bonazzi2023} & Sensors'23 & $112{\times}112$ & 0.43 & 14.8 & 0.76 & 0.12 $\pm$ 0.02 \\
    \midrule
    \multicolumn{7}{@{}l}{\textit{Anomaly detection}} \\
    TinyGLASS~\citep{Bonazzi2026b} & AICAS'26 & $256{\times}256$ & 2.88 & 1\,911.9 & 9.57 & 1.97 $\pm$ 0.06 \\
    \midrule
    \multicolumn{7}{@{}l}{\textit{Pose estimation}} \\
    HigherHRNet & Sony~\citep{imx500models} & $384{\times}288$ & -- & -- & 53.12 & 10.07 $\pm$ 0.13 \\
    \bottomrule
  \end{tabular}}
  \ifdim\wd\imxtablebox>\columnwidth
    \resizebox{\columnwidth}{!}{\usebox{\imxtablebox}}%
  \else
    \usebox{\imxtablebox}%
  \fi
\end{table}

Table~\ref{tab:energy_supp} lists the energy values of Fig.~4 of the main paper.

\begin{table*}[!h]
  \centering
  \caption{Energy per inference $E$, energy above idle $\Delta E$ (mJ) and energy efficiency (MMAC/J, on $\Delta E$) of the six devices with measured energy, the values of Fig.~4 of the main paper.
           Each device is measured on its own supply domain (main paper, Section~3.3).
           ``---'': not deployable (STM32V8: EfficientNet does not fit in memory).
           ``--'': not reported; the IMX500 has no total $E$ (Fig.~4 shows its $\Delta E$).
           Device type: \cpusq MCU without NPU, \npusq MCU with NPU, \dspsq DSP.}
  \label{tab:energy_supp}
  \footnotesize
  \setlength{\tabcolsep}{3pt}
  \resizebox{\textwidth}{!}{%
  \begin{tabular}{@{}l *{4}{rrr}@{}}
    \toprule
    & \multicolumn{3}{c}{\devsq{mobilenetcolor}\textbf{MobileNetV3}} & \multicolumn{3}{c}{\devsq{mcunetcolor}\textbf{MCUNetV1}} & \multicolumn{3}{c}{\devsq{efficientnetcolor}\textbf{EfficientNet}} & \multicolumn{3}{c}{\devsq{squeezenetcolor}\textbf{SqueezeNet}} \\
    \cmidrule(lr){2-4} \cmidrule(lr){5-7} \cmidrule(lr){8-10} \cmidrule(l){11-13}
    \textbf{Device} & $E$ & $\Delta E$ & MMAC/J & $E$ & $\Delta E$ & MMAC/J & $E$ & $\Delta E$ & MMAC/J & $E$ & $\Delta E$ & MMAC/J \\
    \midrule
    \cpusq Xtensa LX7 & 8.45 & 3.57 & 1\,714 & 20.03 & 8.73 & 2\,338 & 63.54 & 26.39 & 1\,219 & 35.12 & 15.46 & 3\,338 \\
    \rowrule
    \cpusq STM32V8 & 4.77 & 1.85 & 3\,305 & 7.58 & 3.45 & 5\,922 & --- & --- & --- & 8.20 & 3.98 & 12\,981 \\
    \rowrule
    \npusq STM32N6 & 1.12 & 0.31 & 19\,478 & 1.64 & 0.50 & 40\,885 & 3.20 & 1.05 & 30\,738 & 0.93 & 0.39 & 132\,232 \\
    \rowrule
    \npusq GAP9 & 0.18 & 0.11 & 57\,736 & 0.44 & 0.29 & 71\,364 & 1.71 & 1.23 & 26\,133 & 0.19 & 0.13 & 393\,969 \\
    \rowrule
    \npusq Alif E7 & 0.55 & 0.12 & 49\,275 & 0.65 & 0.16 & 124\,300 & 1.79 & 0.45 & 72\,162 & 0.36 & 0.12 & 444\,531 \\
    \rowrule
    \dspsq IMX500 & -- & 0.24 & 25\,352 & -- & 0.24 & 86\,483 & -- & 0.35 & 91\,522 & -- & 0.13 & 401\,010 \\
    \bottomrule
  \end{tabular}%
  }
\end{table*}

\end{document}